\documentstyle[pre,aps,multicol,epsf]{revtex}
\begin{document}
\draft
\title{Elasticity of Gaussian and nearly-Gaussian phantom 
networks}
\author{Oded Farago and Yacov Kantor}
\address{School of Physics and Astronomy, Tel Aviv University, Tel
  Aviv 69 978, 
Israel}
\maketitle
\begin{abstract}
We study the elastic properties of phantom networks of Gaussian and
nearly-Gaussian springs. We show that the stress tensor of a Gaussian
network coincides with the conductivity tensor of an equivalent
resistor network, while its elastic constants vanish. We use a
perturbation theory to analyze the elastic behavior of networks of
slightly non-Gaussian springs. We show that the elastic constants of
phantom percolation networks of nearly-Gaussian springs have a power
low dependence on the distance of the system from the percolation
threshold, and derive bounds on the exponents. 

\end{abstract}
\begin{multicols}{2}
\narrowtext
\section{Introduction}
Rubber and gels are large polymeric solid networks formed when
polymers or monomers in fluid solutions are randomly cross-linked by
permanent bonds. This process is called vulcanization or gelation,
when the latter term usually applies to cross-linking of monomers or
very short polymers --- gels; while the former term usually describes
formation of dense networks of long polymers --- rubber. Rubber and
gels are much more flexible than ordinary crystalline solids and,
moreover, may remain in the linear elastic regime even in response to
deformations increasing their dimensions far beyond their original,
unstrained, size. Such a behavior is attributed to the network
structure of these materials, and to the fact that the elastic
restoring forces are of entropic, rather than energetic, origin. The
simplest theory of {\em rubber elasticity}\/ which captures these
essential physical features, is the ``phantom network'' (PN) model
introduced by James and Guth \cite{james}. This model assumes that the
configurations of the different polymer chains are independent of each
other, and neglects the excluded volume interactions between the
monomers. With these simplifying assumptions one can
treat each polymer chain in the network as an ideal one. By averaging
over the positions of the monomers one finds that the probability
density of finding chain ends separated by $\vec{r}$, takes a Gaussian
form $\sim\exp\left[-\frac{1}{2}Br^2\right]$, where $B$ usually
depends on the temperature $T$. The free energy of the chain is
proportional to (minus) the logarithm of this probability density and,
therefore, proportional to $r^2$, as if it is a linear spring of
vanishing equilibrium length, which will be called {\em Gaussian
  spring}. In the PN model, the thermal averages of some quantities can
be calculated analytically due to the Gaussian form of the statistical
weights \cite{weiner}, and this makes it an excellent starting point
for models with excluded volume interactions and
entanglements \cite{deam}.  
 
The problem of {\em gel elasticity}\/ introduces an additional
complication already at the level of the PN model. In gels the network
strands are very short and do not necessarily resemble Gaussian
springs. Nevertheless, one may still construct a Gaussian model of gel
elasticity, simply by replacing each bond of the gel by a Gaussian
spring. In the absence of excluded volume interactions, the validity
of this model is justified by the fact that even if the elementary
pair potential between bonded atoms is very different from that of a
Gaussian spring, the {\em effective}\/ interaction between somewhat more
distant atoms is, almost always, quadratic. This is a well known
feature of long polymer chains~\cite{dgbook}, but it has also been
demonstrated for more complicated networks~\cite{various}. De Gennes
used an analogy between elasticity of networks of Gaussian springs and
conductivity of random resistor networks \cite{degennes}, and argued
that rigidity, just like conductivity, appears at the connectivity 
threshold, when a macroscopically large network spans the system. He
further argued that at the phase transition the shear
modulus and the conductivity should have the same dependence on the
distance of the system from the connectivity threshold. Surprisingly,
the details of the argument of de Gennes have never been worked out,
i.e., there is no detailed calculation of the quantities
characterizing the elastic response of  Gaussian networks, namely the
stress and elastic constants tensors. [There are several analytical
studies of the statistical properties (including the elastic
properties) of systems of Gaussian springs \cite{weiner,kantor}, but
none of them makes such an explicit calculation.] In section
\ref{gaussnet} of this paper we derive exact results for the stress
and elastic constants of Gaussian networks. We prove that the stress
{\em tensor}\/ of a Gaussian elastic network is {\em equal}\/ to the
conductivity {\em tensor}\/ of an equivalent resistor network. A
detailed proof of this equality, which holds for a Gaussian network of
arbitrary topology, is given in the appendix of the paper. We also
show  that the elastic constants of a system consisting of a single
spanning cluster of Gaussian springs {\em vanish}. We discuss the effect
of the finite clusters which model the small molecules formed in the
process of crosslinking and show that they play a crucial role in
stabilizing the system.

In section \ref{nearly} we investigate the elastic behavior of phantom
networks of nearly-Gaussian springs, whose energy dependence on their
extension includes a small quartic term additional to the quadratic
one. A perturbative analysis yields an expression for the elastic
constants. In section \ref{percolation} we use this expression to
evaluate the elastic constants of phantom percolation networks
\cite{stauffer}, close to the percolation threshold $p_c$. We
conjecture a universal scaling law for the elastic constants,
$C\sim(p-p_c)^g$, and derive exact bounds for the scaling exponent
$g$. Section \ref{summary} includes a short summary and discussion of
the main results.    

\section{Elasticity of systems of Gaussian springs---Exact Results}
\label{gaussnet}
\subsection{Definitions in the theory of elasticity}

The theory of elasticity describes deformations of thermodynamic
systems in response to external forces. At a finite temperature, it is
convenient to consider {\em homogeneous}\/ deformations of the {\em
  boundaries}\/ of the system, which can be described by a {\em
  constant}\/ matrix, $M_{ij}$. When the system is strained, the
separation between a pair of surface points, which prior to the
deformation was $\vec{R}$, changes to 
\begin{equation}
r_i=M_{ij}R_j,
\label{trans}
\end{equation}
where the subscripts denote Cartesian coordinates, and summation over
repeated indices is implied. Usually the energy of the system depends
on the relative distances between the atoms. The squared distance
in the deformed system is equal to
\begin{eqnarray}
r^2&=&r_kr_k=M_{ki}R_iM_{kj}R_j \nonumber \\
&=&\left(M^tM\right)_{ij}R_iR_j
\equiv\left(\delta_{ij}+2\eta_{ij}\right)R_iR_j,
\label{rsquare}
\end{eqnarray}
where $M^t$ is the transpose of $M$, and $\eta_{ij}$ is the {\em
  strain}\/ tensor, while $\delta_{ij}$ is the Kr\"{o}necker delta. 
The strain tensor vanishes at the undeformed reference
state. Expanding the mean free energy density in the strain variables  
\begin{equation}
f(\{\eta\})=f(\{0\})+\sigma_{ij}\eta_{ij}
+{1\over 2}C_{ijkl}\eta_{ij} \eta_{kl}+\ldots\ ,
\label{expan}
\end{equation}
we identify the coefficients $\sigma_{ij}$ as the components of the
{\em stress}\/ tensor, while $C_{ijkl}$ are the {\em elastic
  constants} (sometimes referred to as the {\em elastic stiffness
  tensor}\/).  

The elastic constants of a thermodynamic system are related to each
other through certain equalities. The actual number of independent
elastic constants depends on the symmetries of the system. Isotropic
systems, for instance, have only three {\em different}\/ non-vanishing
elastic constants: $C_{xxxx}=C_{yyyy}=C_{zzzz}\equiv C_{11}\,$;
$C_{xxyy}=C_{yyzz}=C_{zzxx}=\ldots\equiv C_{12}\,$; and
$C_{xyxy}=C_{yzyz}=C_{zxzx}=\ldots\equiv C_{44}\,$. Moreover, these
three elastic constants obey an additional relation \cite{wallace}:
$C_{11}-C_{12}=2C_{44}$, which reduces the number of independent
elastic constants of isotropic systems to two. Frequently, one finds
it more useful to describe the elastic behavior in such systems in
terms of the {\em shear}\/ modulus, $\mu$, and the {\em bulk}\/
modulus, $\kappa$, defined by \cite{remark}  
\begin{equation}
\mu=C_{44}-P,
\label{shear}
\end{equation}
and
\begin{equation}
\kappa=\left\{
\begin{array}{ll}
\frac{1}{2}(C_{11}+C_{12}), &  
{\rm for\ 2-dimensional\ systems} \\
\frac{1}{3}(C_{11}+2C_{12}+P), &  
{\rm for\ 3-dimensional\ systems} \\
\end{array}
\right. 
\label{kappa}
\end{equation}
where $P=-\sigma_{xx}=-\sigma_{yy}=-\sigma_{zz}$ is the pressure. When
$\kappa$ and $\mu$ are positive, the system is mechanically stable
\cite{zhoujoos}.  

\subsection{Description of the system}

\begin{figure}[htb]

\epsfysize=15\baselineskip
\centerline{\hbox{
      \epsffile{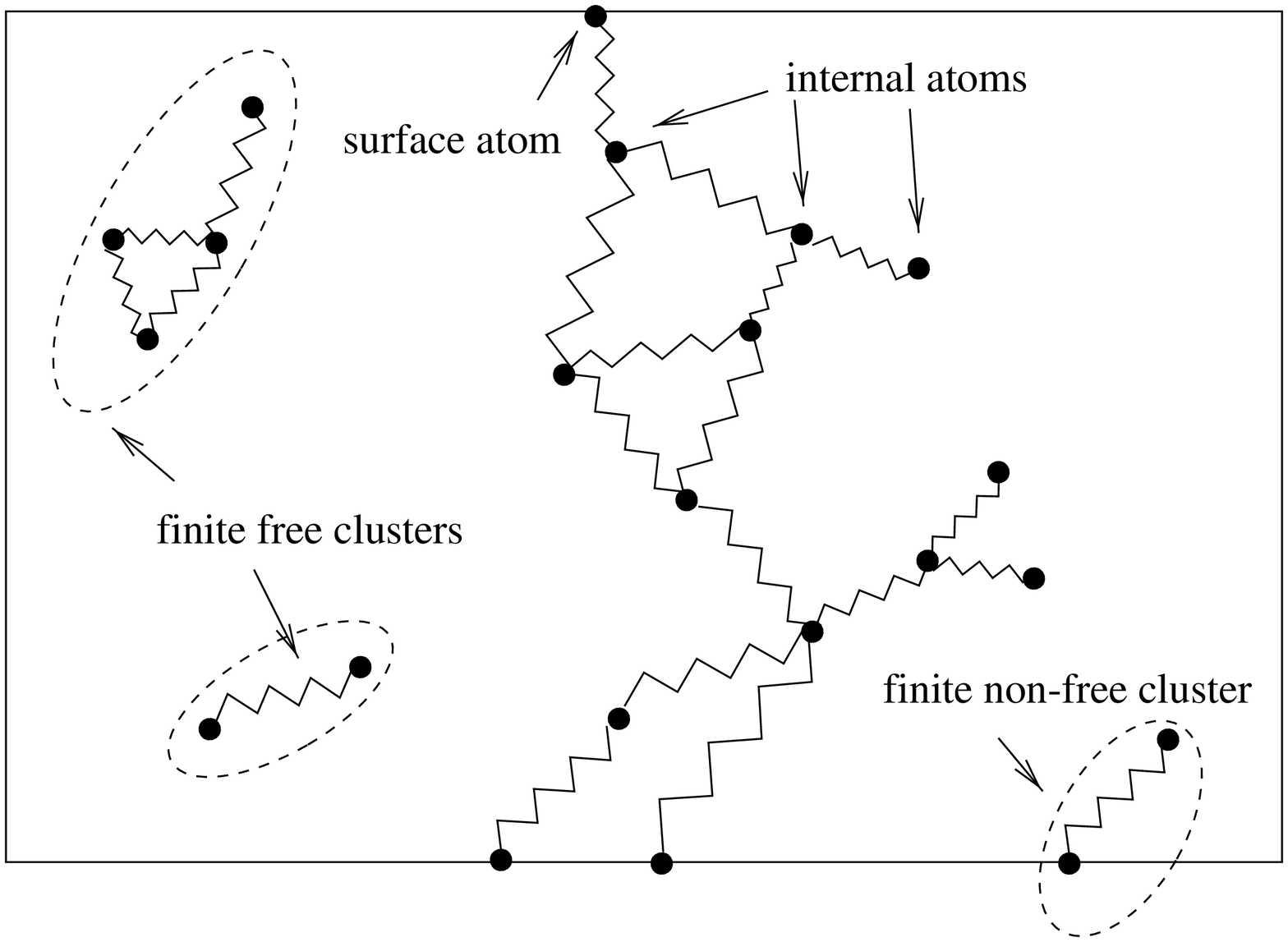}  }}

\caption {\protect A schematic picture of a network of springs. 
  The system includes a spanning elastic network as well as
  some finite clusters. Atoms can be either internal, i.e., free to
  move inside the volume, or external, i.e., attached to a permanent
  positions on the boundaries. Non-free clusters have at least one
  external atom.}
\label{schematic}
\end{figure}

We consider a $d$-dimensional system shown schematically in
Fig.~\ref{schematic}. The black circles in Fig.~\ref{schematic}
represent atoms while the zigzag lines indicate the bonds, attractive
pair-potentials, which connect them in a certain fixed (quenched)
topology. Atoms which are found inside the volume of the systems are
called {\em internal}\/ atoms. {\em Surface}\/ atoms have fixed
coordinates on the boundaries of the system. The bonds connect atoms
into clusters. Clusters containing only internal atoms are {\em
  free}\/ to move in the entire volume. Cluster with both internal and
surface atoms are non-free. Among them, one (and, in some cases,
several) may extend from one side of the system to the opposite
side. This is the ``spanning'' cluster.   

The system which we study in this section consists of point-like atoms
connected by Gaussian springs. The energy of each Gaussian spring is
given by  
\begin{equation} 
\phi_{\alpha\beta}\left(\vec{R}^{\alpha}-\vec{R}^{\beta}\right)
=\frac{1}{2}K^{\alpha\beta}
\left(\vec{R}^{\alpha}-\vec{R}^{\beta}\right)^2=
\frac{1}{2}K^{\alpha\beta}\left(R^{\alpha\beta}\right)^2,
\label{gausspot}
\end{equation}
where $\vec{R}^{\alpha}$ and $\vec{R}^{\beta}$ denote the positions of
atoms $\alpha$ and $\beta$, and $R^{\alpha\beta}$ is the distance
between these atoms. The spring constant $K^{\alpha\beta}$ is assumed
to have a fixed, {\em temperature-independent}, value. The total
elastic energy is given by the sum over the energies of all the
springs
\begin{eqnarray}
E=\sum_{\langle\alpha\beta\rangle}\phi_{\alpha\beta}
=\sum_{\langle\alpha\beta\rangle}
\frac{1}{2}K^{\alpha\beta}\left(R^{\alpha\beta}\right)^2.
\nonumber
\end{eqnarray}

\subsection{Elasticity of the system}

The components of the stress tensor of our system are related to the
pair-potentials, 
$\phi_{\alpha\beta}(R^{\alpha\beta})$, via relation
\begin{equation}
\sigma_{ij}=\frac{1}{V}
\left\langle\sum_{\langle\alpha\beta\rangle}\phi'_{\alpha\beta}
\left(R^{\alpha\beta}\right)
\frac{R_i^{\alpha\beta}R_j^{\alpha\beta}}{R^{\alpha\beta}}
\right\rangle-\frac{NkT\delta_{ij}}{V},
\label{strshh}
\end{equation}
which has been derived thirty years ago by Squire, Holt and Hoover
\cite{shh} as an extension of the Born and Huang theory of elasticity
\cite{bh}, to systems at finite-temperature. In expression
(\ref{strshh}) summation over all distinct pairs of atoms,
$\alpha\beta$, is performed, where $R_i^{\alpha\beta}$ and
$R_j^{\alpha\beta}$ are the $i$th and the $j$th Cartesian components
of $\vec{R}^{\alpha\beta}\equiv\vec{R}^{\alpha}-\vec{R}^{\beta}$. The
symbol $\langle\ \rangle$ indicates a thermal average, while $N$ and
$V$ denote the number of internal atoms and the volume of the system,
respectively. For potential (\ref{gausspot}) the expression
(\ref{strshh}) reduces to 
\begin{equation}
\sigma_{ij}=\frac{1}{V}\left\langle\sum_{\langle\alpha\beta\rangle}
K^{\alpha\beta}
R^{\alpha\beta}_iR^{\alpha\beta}_j\right\rangle
-\frac{NkT}{V}\delta_{ij},
\label{fluct}
\end{equation}  
where the sum is over the connected pairs. 

The two terms in the expression (\ref{fluct}) are called the
configurational and kinetic terms, respectively. The configurational
term can be divided into terms, each one including the sum over the
bonds of one distinct cluster. Since there are no excluded volume
interactions, these terms are independent of each other (the clusters
do not interact with each other), and the contributions of the
different clusters to the stress are additive. We identify the stress
applied by each cluster as 
\begin{equation}  
\sigma_{ij}^{\rm cluster}=\frac{1}{V}\left\langle\sum_
{\langle\alpha\beta\rangle\ \in{\rm \ cluster}}
K^{\alpha\beta}R^{\alpha\beta}_iR^{\alpha\beta}_j\right\rangle
-\frac{N_IkT}{V}\delta_{ij},
\label{clustress}
\end{equation}
where $N_I$ is the number of internal atoms of the cluster. 

\subsection{The contribution of the free clusters}
{\em The gas of free clusters is an ideal gas.}\/ 
Since the different clusters do not ``feel'' each other, it is
intuitively clear that the contribution to the stress of each free
cluster (fc) should be as of a point-like atom. To prove this result
(which is general and does not depend on the particular form of the
pair-potential), we use the fact that for a free cluster (fc), one can
integrate out $d$ degrees of freedom (of say, $\vec{R}^1$) in
Eq.(\ref{clustress}), and express the terms appearing in it in the
relative coordinates $\tilde{R}^{\alpha}_i=R^{\alpha}_i-R^1_i\
\{\alpha=2,\ldots,N_I\}$. (This statement is correct only in the
thermodynamic limit, when the linear size of the system becomes much
larger than the radius of gyration of the free cluster.) One can
easily verify that in the relative coordinates Eq.(\ref{clustress})
may also be written in the following way
\begin{eqnarray} 
\nonumber
\sigma_{ij}=\frac{1}{V}\left\langle\sum_{\alpha=2}^{N_I}
\tilde{R}^{\alpha}_i\frac{\partial{E}}{\partial{\tilde{R}^{\alpha}_j}}
\right\rangle-\frac{N_IkT}{V}\delta_{ij},
\end{eqnarray} 
which from the equipartition theorem gives
$\sigma_{ij}=-(kT/V)\delta_{ij}$. The stress applied by {\em all}\/
the free clusters is simply  
\begin{equation}
\sigma^{\rm fc}_{ij}=-\frac{N_0kT}{V}\delta_{ij}, 
\label{stressfree}
\end{equation}
where $N_0$ is the total number of free clusters. Similarly, the
contribution of the free clusters to the elastic constants is also as
of an ideal gas, given by the kinetic term 
\cite{shh} 
\begin{equation}
C_{ijkl}^{\rm fc}=\frac{2N_0kT}{V}\delta_{il}\delta_{jk}.
\label{elasticfree}
\end{equation}

\subsection{Elasticity of the spanning cluster}

{\em The stress and elastic constants of the spanning network of
  Gaussian springs with temperature-independent force constants, are
  temperature-independent.}\/ 
The free energy $F$ of the spanning network is a function of the
temperature $T$ and the positions of the surface atoms
$\{\vec{R}^s\}$. If the values of these variables change
quasi-statically, then 
\begin{equation}
dF=-SdT+\sum_s\vec{f}^s_{\rm ext}\cdot d\vec{R}^s,
\label{df}
\end{equation}
where $S$ is the entropy, $\vec{f}^s_{\rm ext}$ is the external force
which drags the surface atom $s$ and summation is made over all the
surface atoms. In a quasi-static process, the force
$\vec{f}^s_{\rm ext}$ is balanced by the force $\vec{f}^s$ applied by
the network on atom $s$, namely  
\begin{equation}
-\vec{f}^s_{\rm ext}=\vec{f}^s=\left\langle\sum_{\alpha}K^{\alpha
  s}\left(\vec{R}^{\alpha}-\vec{R}^s\right)\right\rangle,
\label{fs}
\end{equation}   
where summation is over all atoms $\alpha$ connected to atom
$s$. The terms appearing in the thermal average in Eq.(\ref{fs}) are
linear in the coordinates $\vec{R}^{\alpha}$. Since the Boltzmann 
weight is a Gaussian, i.e., an exponent of a quadratic form of the
coordinates, these averages coincide with the most probable values,
namely their values at the energetic ground state, and therefore do
not depend on the temperature. We thus  conclude that $\vec{f}^s$ is a
temperature-independent quantity, and from Eqs.(\ref{df}) and
(\ref{fs}) we readily find that 
\begin{eqnarray}
\frac{\partial ^2 F}{\partial T \partial \vec{R}^s}=
-\frac{\partial \vec{f}^s}{\partial T}=0.
\nonumber
\end{eqnarray}
The last result implies that $F$ can be decomposed into two parts
\begin{eqnarray}
F(T,\{\vec{R}^s\})=F_1(T)+F_2(\{\vec{R}^s\}).
\nonumber
\end{eqnarray}
If we consider homogeneous deformations we may define a reference
system and use the strain variables $\{\eta_{ij}\}$, instead of
$\{\vec{R}^s\}$ 
\begin{eqnarray}
F=F_1(T)+F_2(\{\eta_{ij}\}).
\nonumber
\end{eqnarray} 
The stress and elastic constants are the coefficients in the
$\{\eta\}$-expansion of $F_2$ [see Eq.(\ref{expan})]. Therefore, they
do not depend on the temperature.  

{\em The stress applied by the spanning  network is equal to the
  conductivity of a resistor network with the same topology.}\/   
The stress of the spanning cluster (spc) [Eq.(\ref{clustress})]
\begin{eqnarray}  
\sigma_{ij}^{\rm spc}=\frac{1}{V}\left\langle\sum_
{\langle\alpha\beta\rangle\ \in{\rm \ spc}}
K^{\alpha\beta}R^{\alpha\beta}_iR^{\alpha\beta}_j\right\rangle
-\frac{N_IkT}{V}\delta_{ij},
\nonumber
\end{eqnarray}
can be rewritten in the form
\begin{eqnarray}
\sigma_{ij}^{\rm spc
  }&=&\frac{1}{V}\left\{\left\langle\sum_{\alpha=1}^{N_I}  
R^{\alpha}_i\frac{\partial{E}}{\partial{R^{\alpha}_j}}\right\rangle
+\left\langle\sum_{\langle\alpha s\rangle\ 
\in{\rm \ spc}}
K^{\alpha s}R^s_iR^{s\alpha}_j\right\rangle\right\}\nonumber\\
&-&\frac{N_IkT}{V}\delta_{ij},
\label{infstress2}
\end{eqnarray} 
where the first sum is over all the internal atoms, while the second
sum is over all the bonds connecting internal and surface atoms. (The
subscripts $s$ and $\alpha$ denote surface and internal atoms,
respectively.) In the thermodynamic limit we deduce from the
equipartition theorem that the first and the third (kinetic) terms in
Eq.(\ref{infstress2}) cancel each other. We are thus left only with
the second term 
\begin{equation}
\sigma_{ij}^{\rm spc}=\frac{1}{V}\left[
\sum_{\langle\alpha s\rangle\ \in{\rm \ spc}}
K^{\alpha s}R^s_i\left\langle R^{s\alpha}_j\right\rangle\right].
\label{infstress3}
\end{equation}

The thermal averages in Eq.(\ref{infstress3}) are of quantities which
are linear in the coordinates of the internal atoms and therefore may
be replaced by the equilibrium values of these quantities (see earlier
in this section). The equilibrium values of $\vec{R}^{\alpha}$
minimize the energy of the spanning cluster 
\begin{eqnarray}
E^{\rm spc}&=&\sum_{\langle\alpha \beta\rangle\ \in{\rm \
    spc}}\frac{1}{2}
K^{\alpha\beta}\left(\vec{R}^{\alpha}-\vec{R}^{\beta}\right)^2
\nonumber\\
&=&
\sum_{j=1}^d\left[\sum_{\langle\alpha\beta\rangle\ \in  
{\rm \ spc}}\frac{1}{2}
K^{\alpha\beta}\left(R^{\alpha}_j-R^{\beta}_j\right)^2\right]
\nonumber\\
&\equiv&
\sum_{j=1}^d E^{\rm\ spc}_j. 
\label{espc}
\end{eqnarray}
The dependence of $E^{\rm spc}$ on the components $R^{\alpha}_j$
corresponding to one Cartesian direction, $j$, is included in the term
$E^{\rm spc}_j$. The problem of finding the equilibrium values of
$\vec{R}^{\alpha}$ decouples into $d$ scalar problems of finding the
equilibrium values of $R^{\alpha}_j$. In order to calculate these
values we need to solve $d$ sets of linear equation (one set for each
Cartesian component):   
\begin{equation}
\sum_{\beta}
K^{\alpha\beta}\left(R^{\alpha}_j-R^{\beta}_j\right)=0,
\label{forceq}
\end{equation}
corresponding to the vanishing of the $j$th component of the force
acting on each internal atom. (For each atom $\alpha$, summation in
the relevant equation is over all atoms $\beta$ connected to it.) 

Let us define a resistor network with the same connectivity as the
elastic network, in which each spring is replaced by a resistor with
conductance $K^{\alpha\beta}$. The values of the electric potential at
the internal nodes, $\{\varphi^{\alpha}\}$, are obtained by
minimization of the heat power produced in the network,
$P=\sum_{\langle\alpha\beta\rangle}
K^{\alpha\beta}(\varphi^{\alpha}-\varphi^{\beta})^2$. Except for a
prefactor of $\frac{1}{2}$, $P$ is identical with $E^{\rm\ spc}_j$
(\ref{espc}), where $\varphi^{\alpha}$ plays the role of
$R^{\alpha}_j$. If we replace $R^{\alpha}_j$ by $\varphi^{\alpha}$ in
the force equations (\ref{forceq}), we obtain the set of Kirchoff
equations enforcing the vanishing of the sum of currents entering the
internal nodes of the network. By replacing $R^{\alpha}_j$ by
$\varphi^{\alpha}$, we define a mapping of the mechanical problem to
an electrostatic one. In fact, we have $d$ different electrostatic
problems corresponding to each Cartesian component of the mechanical
problem. They differ from each other in their boundary conditions,
namely the values of the electric potential on the surface nodes,
$\{\varphi^s\}$. In the $j$th electrostatic problem, we set
$\varphi^s$ equal to $R^s_j$, i.e., we assume that the electric
potential at each boundary point is equal to the $j$th Cartesian
coordinate of the point.   

The interesting question now is what is the analog of the stress
tensor in the electrostatic problem. This appears to be the
conductivity tensor, $\Sigma_{ij}$ defined by 
\begin{eqnarray}
\langle j_i\rangle=\Sigma_{ij}\langle E_j\rangle,
\nonumber
\end{eqnarray}
where $\langle\vec{j}\rangle$ and $\langle\vec{E}\rangle$ are the {\em
  volume averages}\/ of the current density and the electric field,
respectively. More precisely, if we follow the mapping defined above
we have the {\em exact}\/ equality  
\begin{equation}
\sigma_{ij}=\Sigma_{ij}.
\label{identity}
\end{equation} 
A detailed proof of this equality is given in the appendix to this
paper. Here we just note that the proof consists of two steps: In the
first step we show that in the $j$th electrostatic problem, because
of the choice of boundary conditions, $\langle\vec{E}\rangle$ is a
unity electric field pointing in the $(-j)$th direction. In the
presence of such an electric field $\langle
j_i\rangle=-\Sigma_{ij}$. On the next step of the proof we show that
$-\langle j_i\rangle$, and therefore $\Sigma_{ij}$, are given by the 
electrostatic equivalent of Eq.(\ref{infstress3})
\begin{equation}
\Sigma_{ij}^{\rm spc}=\frac{1}{V}\left[
\sum_{\langle\alpha s\rangle\ \in{\rm \ spc}}
K^{\alpha s}R^s_i\left(\varphi^s-\varphi^{\alpha}\right)\right],
\label{conductivity}
\end{equation}
and therefore Eq.(\ref{identity}) is valid. 

{\em The elastic constants of the spanning network vanish.}\/
We have already shown that $C_{ijkl}$, the elastic constants of the
spanning cluster of Gaussian springs with temperature-independent
force constants, are temperature-independent. Therefore, we may
calculate them at any temperature, and in particular at $T=0$. At zero
temperature the free energy coincides with the internal energy, given
by Eq.(\ref{espc}), where $\{\vec{R}^{\alpha}\}$, the positions of the
internal nodes, take their equilibrium values. Suppose now that the
system is homogeneously strained. The positions of the surface nodes,
$\{\vec{R}^s\}$, change according to the linear transformation
(\ref{trans}), with a constant matrix $M_{ij}$. [Transformation
(\ref{trans}) was originally defined for the separation between
surface points. However, we can always set the origin of axes to be on
the original (unstrained) surface, and in this case the transformation
applies to the positions of the surface points.] In order to find the
new equilibrium positions of the internal atoms, in the strained
system, we need to solve the set of equation (\ref{forceq}) with the
new boundary conditions. Since both the equations and the
transformation of the boundary conditions are linear, the new solution
is given by $r^{\alpha}_i=M_{ij}R^{\alpha}_j$. The elastic energy of
the strained spanning cluster is given by [see Eqs.(\ref{rsquare}) and
(\ref{espc})]   
\begin{eqnarray}
E^{\rm spc}&=&\frac{1}{2}\sum_{\langle\alpha s\rangle\ \in{\rm \
    spc}}K^{\alpha\beta} (r^{\alpha\beta})^2
\nonumber\\
&=&
\frac{1}{2}\sum_{\langle\alpha s\rangle\ \in{\rm \
    spc}}K^{\alpha\beta}
\left[(M^tM)_{ij}R^{\alpha\beta}_iR^{\alpha\beta}_j\right]
\nonumber\\
&=&\frac{1}{2}\sum_{\langle\alpha s\rangle\ \in{\rm \ spc}}
K^{\alpha\beta}\left[(2\eta_{ij}+\delta_{ij})R^{\alpha\beta}_i
R^{\alpha\beta}_j\right].
\nonumber
\end{eqnarray}
This gives the dependence of $E$ on the strain variables, which
include only linear terms in $\eta_{ij}$. Since the elastic
constants are the coefficients of the quadratic terms in the
$\{\eta\}$-expansion of the free energy [Eq.(\ref{expan})], we
conclude that
\begin{equation}
C^{\rm spc}_{ijkl}\equiv 0.
\label{ceq0}
\end{equation}
\subsection{The stability of systems of Gaussian springs}
 
We have mentioned earlier in this section that stable solid
thermodynamic systems have positive bulk and shear moduli, $\kappa$
and $\mu$ [Eqs.(\ref{shear}) and (\ref{kappa})]. In phantom systems,
the contributions of the spanning cluster and the ensemble of free
clusters to $\kappa$ and $\mu$ are additive. Due to the vanishing of
the elastic constants of the spanning cluster (\ref{ceq0}), we find
that its contribution to the elastic moduli is: $\mu^{\rm spc}=-P^{\rm
  \,spc}>0$, and $\kappa^{\rm spc}=0$ (two-dimensions) or $\kappa^{\rm
  spc}=P^{\rm \,spc}/3<0$ (three-dimensions) [$P^{\rm \,spc}$ is the
negative (stretching) pressure applied by the spanning cluster]. The
fact that $\kappa$ is not positive means that the spanning cluster
alone is not stable against homogeneous volume fluctuations. The
contribution of the free clusters to the elastic moduli is as of an
ideal gas, given by: $\mu^{\rm fc}=0$, and $\kappa^{\rm fc}=N_0kT/V$
[see Eqs.(\ref{shear}), (\ref{kappa}), (\ref{stressfree}) and
(\ref{elasticfree})]. The vanishing of the of the shear modulus simply
indicates that the collection of free clusters is a fluid. The
positive contribution of the free clusters to the bulk modulus is
crucial for the stability of the system. Two-dimensional Gaussian
networks are stabilized in the presence of free clusters since
$\kappa=\kappa^{\rm spc}+\kappa^{\rm fc}=\kappa^{\rm
  fc}>0$. Three-dimensional systems are stabilized provided that the
positive contribution of the free clusters to $\kappa$ overcomes the
negative contribution of the spanning cluster. 

\section{Elasticity of systems of nearly-Gaussian springs}
\label{nearly}

The elastic response of polymers and polymeric networks is as of
systems of Gaussian springs only in the first approximation. It always
includes a non-linear part, which becomes significant when the network
is sufficiently stretched, much beyond its characteristic thermal
lengths \cite{james,fisherman}. In order to study the nature of this
correction, we consider networks of springs having the spring energies
 \begin{equation}
\phi_{\alpha\beta}\left(R^{\alpha\beta}\right)
=\frac{1}{2}K^{\alpha\beta}\left(R^{\alpha\beta}\right)^2+
\frac{1}{4}a^{\alpha\beta}\left(R^{\alpha\beta}\right)^4.
\label{nearlypoten}
\end{equation}
Our choice for the spring energy is inspired by the free energy of a
finite long polymer chain \cite{james}, where the leading correction
to the linear relation between the force and the chain end-to-end
vector $\vec{f}=K\vec{R}$, is a term proportional to $R^2\vec{R}$. The
elastic energy of the system is given, again, as the sum of all
springs energies    
\begin{eqnarray}
E=\sum_{\langle\alpha\beta\rangle}\phi_{\alpha\beta}
&=&\sum_{\langle\alpha\beta\rangle}\left[
\frac{1}{2}K^{\alpha\beta}\left(R^{\alpha\beta}\right)^2
+\frac{1}{4}a^{\alpha\beta}\left(R^{\alpha\beta}\right)^4\right]
\nonumber\\
&\equiv& E_0+E_1.
\label{energy2}
\end{eqnarray}
We assume that $E_1\ll E_0$, and treat the quartic term
perturbatively. In fact, we will make a more restrictive assumption
that for each bond $a^{\alpha\beta}\left(R^{\alpha\beta}\right)^4\ll
K^{\alpha\beta}\left(R^{\alpha\beta}\right)^2$. Since the quadratic
term $E_0$ does not make any contribution to the elastic constants, we
will mainly focus on the contribution of the perturbation term, $E_1$,
to them.  

{\em In the lowest order of a perturbation theory, the elastic
  constants of the network are temperature independent.}\/  
Substituting the pair potential (\ref{energy2}) into expression
(\ref{strshh}) for the stress tensor, and expanding this expression to
the first order in $a^{\alpha\beta}$, yields
\begin{eqnarray}
\sigma_{ij}=\sigma_{ij}^0&+&\frac{1}{V}
\left\langle\sum_{\langle\alpha\beta\rangle}
a^{\alpha\beta}\left(R^{\alpha\beta}\right)^2R^{\alpha\beta}_i
R^{\alpha\beta}_j\right\rangle_0
\nonumber\\
&-&\frac{1}{VkT}
\left\langle\delta\left(\sum_{\langle\alpha\beta\rangle}
K^{\alpha\beta}R^{\alpha\beta}_iR^{\alpha\beta}_j\right)\delta E_1
\right\rangle_0,
\label{nearlystress}
\end{eqnarray}  
where $\delta A\equiv A-\langle A\rangle_0$ denotes a thermal
fluctuation of the quantity $A$, and $\langle\ \rangle_0$ denotes a
thermal average with the (unperturbed) Gaussian Boltzmann weight
$\exp(-E_0/kT)$. $\sigma_{ij}^0$ is the stress tensor of the
corresponding Gaussian network (where $a^{\alpha\beta}\equiv 0$),
given by Eq.(\ref{fluct}), which can be also expressed by its value at
$T=0$ 
\begin{equation}
\sigma_{ij}^0=\frac{1}{V}
\sum_{\langle\alpha\beta\rangle}\left[K^{\alpha\beta}
\left(R^{\alpha\beta}_0\right)_i\left(R^{\alpha\beta}_0\right)_j
\right].
\label{stresszero}
\end{equation}
In the above expression $\left(R^{\alpha\beta}_0\right)_i$ is the
$i$th Cartesian component of the bond vector
$\vec{R}^{\alpha\beta}_0$, at the ground state of the unperturbed
Gaussian network.   

The next step is to substitute the pair potential (\ref{energy2}) into
the expression for the elastic constants \cite{shh} (see also Eq.(7)
in Ref.\cite{farago}). By expanding this expression to the first order
in $\{a^{\alpha\beta}\}$, and using the fact that for the Gaussian
network $C_{ijkl}\equiv 0$ (\ref{ceq0}), we find that
\begin{eqnarray}
C_{ijkl}=\frac{2}{V}\left\langle\sum_{\langle\alpha\beta\rangle}
a^{\alpha\beta}R^{\alpha\beta}_iR^{\alpha\beta}_jR^{\alpha\beta}_k
R^{\alpha\beta}_l\right\rangle_0+\left\langle X\right\rangle_0, 
\nonumber 
\end{eqnarray}
where $X$ is combination of terms, each of which includes the thermal
fluctuations of some quantities. Since at $T=0$ there are no thermal
fluctuations, that term vanishes and we readily find that 
\begin{eqnarray}
C_{ijkl}(T=0)=\frac{2}{V}\left\langle\sum_{\langle\alpha\beta\rangle}
a^{\alpha\beta}R^{\alpha\beta}_iR^{\alpha\beta}_jR^{\alpha\beta}_k
R^{\alpha\beta}_l\right\rangle_0
\nonumber\\
=\frac{2}{V}
\sum_{\langle\alpha\beta\rangle}\left[
a^{\alpha\beta}\left(R^{\alpha\beta}_0\right)_i
\left(R^{\alpha\beta}_0\right)_j\left(R^{\alpha\beta}_0\right)_k
\left(R^{\alpha\beta}_0\right)_l\right].
\label{cexpression}
\end{eqnarray}
The second equality in the above equation is obtained by equating the
expression inside $\langle\ \rangle_0$ to its value at equilibrium (at
zero temperature the thermal average coincides with this value). 

At a finite temperature we may write the elastic constants as the
product of $C_{ijkl}(T=0)$, and a dimensionless function, which may
depend only on terms of the form
$(kT\,a^{\alpha\beta})/(K^{\gamma\delta}K^{\epsilon\zeta})$. Expanding
the function into power series in these variables yields
\begin{eqnarray} 
&C_{ijkl}&=
\nonumber\\
&C_{ijkl}&(T=0)\left[1+
\left({\rm linear \ terms\ in\ }\left\{\frac{kT\,a^{\alpha\beta}}{
K^{\gamma\delta}K^{\epsilon\zeta}}\right\}\right)+\ldots\right].
\nonumber
\end{eqnarray}
Since $C_{ijkl}(T=0)$ is a linear function in the quantities
$a^{\alpha\beta}$, and since we are interested only in the
first order correction due to the perturbation (namely, in terms
linear in $a^{\alpha\beta}$), we conclude that to the lowest order in
$a^{\alpha\beta}$, $C_{ijkl}$ are temperature independent, and
therefore given by the above expression (\ref{cexpression}).  

\section{Elasticity of Phantom Percolation networks}
\label{percolation}
\subsection{The percolation model}

One of the models which has been proposed to describe the process of
gelation is {\em percolation}\/ \cite{stauffer}. In the percolation
model, the sites or the bonds of a lattice are randomly occupied by,
respectively, atoms or bonds, with an occupation probability
$p$. In the site percolation model, one links every two
neighboring occupied sites, while in the bond percolation model one
assumes that all the sites are occupied by atoms and each pair of
neighbors is linked if the bond between the atoms exists. Within the
percolation model, the gel point is identified with the percolation
threshold, the critical site/bond concentration above which a spanning
cluster is formed. The percolation model predicts that close to the
percolation threshold, $p_c$, quantities like the mean cluster mass,
typical cluster linear size and gel fraction, have power-law
dependence on $(p-p_c)$. The relevant exponents are universal and
depend only on the dimensionality of the system, but not on the
atomic-scale features of the system. The values of these exponents
have been measured experimentally for various gel systems
\cite{adam}. A fairly good agreement have been found between the
measured exponents and their values as predicted by the percolation
model, what proves the applicability of the percolation model to
gelation.   

The situation concerning the elastic behavior of gels is not that
clear. The main question is whether the shear modulus also follows a
scaling law $\mu\sim(p-p_c)^f$ with a universal exponent
$f$. Experimental values of this exponent measured for different
polymeric systems are very scattered \cite{experiments}. On the
theoretical side, it has been demonstrated that at $T=0$, the elastic
behavior of percolation systems depends on the nature of the
interactions in the system. For non-stressed central force networks
the rigidity threshold occurs at a concentration of bonds much larger
than $p_c$ \cite{central}. If bond bending forces are present,
rigidity and percolation thresholds coincide; however the rigidity
exponent $f$, is considerably larger than the conductivity exponent,
$t$, suggesting that the two problems belong to different universality
classes \cite{bending}. As the number of models of elasticity of
random systems increased, it became clear that de Gennes' conjecture
about the identity of the exponent $f$ to the conductivity exponent
$t$ \cite{degennes} can be justified only within models which
``reduce'' the thermodynamic behavior of gels to so called ``scalar
elasticity'' models \cite{alexander}. Recently, the equality $f=t$ was
measured by Plischke {\em et al.}\/~in a numerical study of {\em
  phantom}\/ central force percolation networks at $T\neq 0$
\cite{plischke}. The authors attributed this elastic behavior to the
entropic part of the elastic free energy.  

\subsection{Elasticity of percolation networks}

We would like to apply our results from sections \ref{gaussnet} and
\ref{nearly} to phantom percolation networks of identical springs
having the energy $E=\frac{1}{2}KR^2$ (Gaussian network) or
$E=\frac{1}{2}KR^2+\frac{1}{4}aR^4$ (nearly-Gaussian network). We
discuss the critical elastic behavior of such networks, in the regime
where the correlation length $\xi\sim(p-p_c)^{-\nu}$ is much larger
than the characteristic atomic length scale $b$, but much smaller then
the linear size of the system $L$. The correlation length is the
length scale below which the spanning cluster has a fractal structure
and above which the system is homogeneous. A quantity that follows a
power law $\sim(p-p_c)^Y\sim\xi^{-(Y/\nu)}$ when $L\gg\xi$, scales
as to $L^{-(Y/\nu)}$ when $\xi\gg L$. (At $p_c$ the latter power law
is always relevant because $\xi$ is infinite.) Since $\xi\gg b$, we
expect the structure of the spanning cluster to ``forget'' the details
of the lattice, and have the elastic properties of an isotropic
system. In the Gaussian case, the tensorial equality
$\sigma_{ij}=\Sigma_{ij}$ (\ref{identity}) becomes a scalar equality
$-P=\Sigma$. Also, because of the vanishing of the elastic constants
of Gaussian networks (\ref{ceq0}), we have for the shear modulus of
the spanning cluster that $\mu=C_{44}-P=-P=\Sigma$
(\ref{shear}). Close to the percolation threshold, the conductivity
scales as $\Sigma\sim(p-p_c)^t$, and therefore we conclude that for
Gaussian networks  
\begin{equation}
\mu=-P=\Sigma\sim(p-p_c)^t,
\label{scaling}
\end{equation}
in accordance with de Gennes' argument. This result is not changed if
we also include the finite clusters, since the latter make no
contribution to the shear modulus (just as they do not contribute to
the conductivity of the system). The equality of the shear modulus and
the stress, a signature of Gaussian elasticity, was observed
numerically in Ref.\cite{plischke}. 

In the nearly-Gaussian case, we have from Eq.(\ref{nearlystress})
that the leading term in the expression for the stress is the Gaussian
term, and therefore we expect to have the same scaling behavior as in
Eq.(\ref{scaling}). What distinguishes non-Gaussian networks from
purely Gaussian ones is the non-vanishing elastic constants of the
former. For percolation networks it is reasonable to assume that the
elastic constants also follow a power law $C\sim(p-p_c)^g$. The
elastic constants of a nearly-Gaussian networks should be ``almost''
zero, namely much smaller than the network stress. Therefore, the
perturbative analysis in section \ref{nearly} would be self-consistent
only if it yields that the exponent $g>f$. We can use expression
(\ref{cexpression}) for the elastic constants to derive exact bounds
on the value of the exponent $g$. Consider a percolation network of
linear size $L$ in $d$ dimensions at $p_c$. An upper bound on the
exponent $g$ is obtained by including only a partial set of the bonds
of the spanning cluster in the sum in expression
(\ref{cexpression}). We take the set of singly connected bonds
(SCBs), which are such bonds that removal of each one of them
disconnects the spanning cluster. Their number scales as $L^{1/\nu}$
\cite{coniglio}. The force acting on a SCB is the total force applied
on the surface of the system, which is proportional to $PL^{(d-1)}\sim
L^{(-t/\nu+d-1)}$. The length to which a SCB is stretched, $(R_{\rm
  SCB})_0$, is proportional to the force, and therefore have the same
scaling form 
\begin{equation}
(R_{\rm SCB})_0\sim L^{(-t/\nu+d-1)},
\label{scalerscb}
\end{equation}
and consequently from Eq.(\ref{cexpression}) we get
\begin{eqnarray}
C\sim L^{-g/\nu}\geq L^{-d}L^{1/\nu}L^{4(-t/\nu+d-1)},  
\nonumber
\end{eqnarray}
which yields the upper bound $g\leq(4t-1)-\nu(3d-4)$. A lower bound
for $g$ is obtained by noting that for any bond other than the SCBs,
$(R_{\rm bond})_0<(R_{\rm SCB})_0$. That is because the SCBs are the
only bonds which experience the total force acting on the system. We
use this fact in expression (\ref{cexpression}) and write that  
\begin{eqnarray}
C\sim L^{-g/\nu}\leq\left[(R_{\rm SCB})_0\right]^2\left\{
\frac{1}{V}\sum_{\rm bonds}a[(R_{\rm bond})_0]^2\right\}.
\nonumber
\end{eqnarray}
The term in braces in the above inequality is, however, proportional
to the pressure [see Eq.(\ref{stresszero})], which scales like
$L^{-t/\nu}$. This, together with result (\ref{scalerscb}), bring us
to the lower bound $g\geq 3t-2\nu(d-1)$. Using the known values of the
exponents $t$ and $\nu$ \cite{conductivity,stauffer}, we find that in
three-dimensions $2.48\leq g\leq 2.6$. In six-dimensions both bounds
coincide to give $g=4$. This last result reflects the fact that in
six-dimensions essentially all the bonds of the network are SCBs. In
two-dimensions we have the bounds $1.22\leq g\leq 1.52$. However, we
must mention a special feature of the two-dimensional case which
questions the validity of the ``nearly'' Gaussian model. The model
assumes that the contribution of the quartic term to the spring energy
is small compared to the quadratic term [Eq.(\ref{nearlypoten})]. This
happens only if the bond length satisfies 
\begin{equation}
R_{\rm bond}\ll \left(K/a\right)^{1/2}. 
\label{criterion}
\end{equation}  
The longest bonds in the network are the bonds that {\em inside a cell
  of size $\xi^d$}\/ serve as SCBs. Close to $p_c$, their length
scales like  
\begin{eqnarray}
R_{\rm bond}\sim\xi^{-t/\nu+(d-1)}
\sim(p-p_c)^{t-\nu(d-1)}\equiv(p-p_c)^y.
\nonumber
\end{eqnarray}
In two-dimensions the exponent $y<0$, what implies that the length of
the SCBs diverges, and certainly does not satisfy criterion
(\ref{criterion}). The problem is not limited to the SCBs only, but
is relevant to a larger fraction of the bonds including the
doubly-connected bonds, triply-connected bonds, and so on. It is
difficult to predict, a priori, whether this observation should modify
the results of the nearly-Gaussian model from section
\ref{nearly}. Note that we do not encounter such a problem for
dimensionality larger than two, where the exponent $y$ is positive.    

\section{Summary and Discussion}
\label{summary}

We have studied the elastic properties of phantom Gaussian and
nearly-Gaussian networks. For Gaussian networks, the stress and
elastic constants were calculated exactly. We found that a
characteristic feature of Gaussian networks is the vanishing of their
elastic constants. This feature is both temperature and
network-topology independent. We also proved the equality between the
stress tensor of a Gaussian elastic network to the conductivity tensor
of a resistor network, in which the conductance of each resistor is
equal to the corresponding spring constant $K^{\alpha\beta}$. This
result quantifies the somewhat vague statement about an analogy
between elasticity of Gaussian networks to conductivity of resistors
networks.  

We have investigated the non-linear correction to the elastic
behavior, by studying the properties of networks of springs whose
energies include small quartic terms, additional to the leading
quadratic (Gaussian) terms. While the stress tensor is still dominated
by the contribution of the quadratic term, the elastic constants
(which vanish in the Gaussian network) are solely due to the
non-Gaussian correction. We calculated the elastic constants to the
first order in perturbation theory. 

Finally, we applied the results of both the Gaussian and the
nearly-Gaussian models, to describe the elastic behavior of
phantom percolation networks close to the percolation
threshold. Obviously, the well known result that the shear modulus
follows the same scaling law, $\mu\sim(p-p_c)^t$, like the
conductivity, was recovered. We made a new prediction that the elastic
constants also follow a scaling law $C\sim(p-p_c)^g$, with exponent
$g>t$, and found bounds on the values of the exponent $g$. 

This work was supported by the Israel Science Foundation through Grant
No. 177/99.

\appendix
\section{The Conductivity Tensor of Finite Resistor networks.}
We consider a network whose bonds are resistors of conductance
$K^{\alpha\beta}$, where the superscripts $\alpha$ and $\beta$ label
the nodes which the particular resistor connects. The network is
finite and has an arbitrary topology, i.e., we make no assumption on
the symmetry. We denote by $\vec{R}^{\beta}$ the position of the node
$\beta$ and by $\varphi^{\beta}$ the electric potential at the
node. The network is placed inside a rectangular box of volume
$V=L_1\times L_2\times\ldots\times L_d$, where $L_i$ is the length
of the box along the $i$th Cartesian direction. (The derivation
presented here can be easily generalized to systems of arbitrary
shape.) The nodes of the network which are located on the surface of
the system are called surface nodes, and we label them with the
superscript $s$. The  rest of the nodes are called the internal nodes,
which we denote with the superscript $\alpha$. The superscripts
$\beta$ and $\gamma$ will be used to denote nodes of both types.    

The conductivity of an electrical system is a tensor, $\Sigma_{ij}$,
defined by 
\begin{equation}
\langle j_i\rangle=\Sigma_{ij}\langle E_j\rangle,
\label{definition}
\end{equation}
where the subscripts denote Cartesian coordinates and summation over
repeated indices is implied, while $\langle\vec{j}\rangle$ and
$\langle\vec{E}\rangle$ are the volume averages of the current density
and the electric field, respectively. This definition of $\Sigma_{ij}$
applies to continuous electrical systems. It can be generalized to
discrete networks, if we define the current density by a set of Dirac
$\delta$-functions representing the currents in the bonds. Let us
assume now that the electric potential, $\varphi$, applied on the
surface of the network is such that on each surface point it is equal
to the $j$th Cartesian coordinate of the point. Since
$\vec{E}=-\vec{\nabla}\varphi$, we have
\begin{eqnarray}
\langle E_i\rangle&=&\frac{1}{V}\int E_i\,dV=-\int
\frac{\partial \varphi}{\partial x_i}\,dV\nonumber\\
&=&\frac{1}{V}\left[-\int_{x_i=L_i}\varphi\,dS+\int_{x_i=0}
\varphi\,dS\right],\nonumber
\end{eqnarray}
where the surface integration is over the boundaries $x_i=0$ and
$x_i=L_i$, normal to the $i$th direction. However, with our choice
for the electric potential on the boundaries, $\varphi=x_j$, it is
easy to see that  $\langle E_i\rangle=-\delta_{ij}$, where
$\delta_{ij}$ is the Kr\"{o}necker delta. 

The mean current density $\langle j_i \rangle$, is given by 
\begin{equation}
\langle j_i\rangle=\frac{1}{V}\int j_i\,dV
\label{ji}
\end{equation}
As we have already noted, the above definition (\ref{ji}) applies to
continuous electrical systems. To make it applicable to resistor
networks we need to write the current density as a sum of Dirac
$\delta$-functions representing the currents in the ``linear''
resistors. With this formal representation, the contribution to
$\langle j_i\rangle$ of each resistor is given by the line-integral 
\begin{eqnarray}
\nonumber
\int_{\vec{R}^\alpha}^{\vec{R}^\beta}I^{\alpha\beta}\,dx_i
=K^{\alpha\beta}\left(\varphi^{\alpha}-\varphi^{\beta}\right)
\left(R_i^{\beta}-R_i^{\alpha}\right),
\end{eqnarray}
where
$I^{\alpha\beta}=K^{\alpha\beta}\left(\varphi^{\alpha}
-\varphi^{\beta}\right)$ is the current across the resistor between
nodes $\alpha$ and $\beta$. Adding the contributions of all the
resistors we find that   
\begin{eqnarray}
\nonumber
\langle j_i\rangle=\frac{1}{V}
\sum_{\langle\alpha\beta\rangle}
K^{\alpha\beta}\left(\varphi^{\alpha}-\varphi^{\beta}\right)
\left(R^{\beta}_i-R^{\alpha}_i\right).
\end{eqnarray}
We may write the last result is a slightly different way
\begin{eqnarray}
\langle j_i\rangle&=&\frac{1}{2V}\left\{
\sum_{\gamma}\sum_{\beta}K^{\gamma\beta}\Theta^{\gamma\beta}
\left(\varphi^{\gamma}-\varphi^{\beta}\right)
\left(-R^{\gamma}_i\right)\right.
\nonumber\\
&+&\left.
\sum_{\gamma}\sum_{\beta}K^{\gamma\beta}\Theta^{\gamma\beta}
\left(\varphi^{\gamma}-\varphi^{\beta}\right)
R^{\beta}_i\right\}\nonumber \\
&=&\frac{1}{V}\left\{
\sum_{\gamma}\left(-R^{\gamma}_i\right)
\left[\sum_{\beta}K^{\gamma\beta}\Theta^{\gamma\beta}
\left(\varphi^{\gamma}-\varphi^{\beta}\right)\right]\right\},
\nonumber
\end{eqnarray}
where the variable $\Theta^{\gamma\beta}$ takes the value 1 if the
nodes $\gamma$ and $\beta$ are connected by a resistor and if at least
one of them is an internal node; and the value 0, otherwise. The sums
in square brackets corresponding to internal nodes $\gamma=\alpha$
vanish due to the Kirchoff ``junction rule'' for the vanishing of the
sum of currents entering an internal node:  
\begin{eqnarray}
\nonumber
\sum_{\beta}K^{\alpha\beta}\Theta^{\alpha\beta}
\left(\varphi^{\alpha}-\varphi^{\beta}\right)=0.
\end{eqnarray}
We are left with the contribution of the surface nodes $\gamma=s$
only, i.e.,
\begin{eqnarray}
\nonumber
\langle j_i\rangle=\frac{1}{V}\left\{
\sum_{s}R^{s}_i\sum_{\beta}K^{\beta s}\Theta^{\beta s}
\left(\varphi^{\beta}-\varphi^{s}\right)
\right\}.
\end{eqnarray}
This last result can be also represented by summation over all the
resistors $\langle\alpha s\rangle$,  between surface and internal
nodes
\begin{eqnarray}
\nonumber
\langle j_i\rangle=\frac{1}{V}\left[
\sum_{\langle \alpha s\rangle}K^{\alpha s}R^{s}_i
\left(\varphi^{\alpha}-\varphi^{s}\right)\right].
\end{eqnarray}
Finally, since the electric field is equal to $\langle E_i
\rangle=-\delta_{ij}$, we have from Eq.(\ref{definition}) that 
\begin{eqnarray}
\Sigma_{ij}= -\langle j_i\rangle=\frac{1}{V}\left[
\sum_{\langle \alpha s\rangle}K^{\alpha s}R^{s}_i
\left(\varphi^{s}-\varphi^{\alpha}\right)\right].
\nonumber
\end{eqnarray}
We have obtained expression (\ref{conductivity}), which we constructed
by mapping expression (\ref{infstress3}) for $\sigma_{ij}$ into the
electrostatic problem. This proves that indeed
$\sigma_{ij}=\Sigma_{ij}$. Note that $\Sigma_{ij}$ does not depend on
the positions of the internal nodes but only on the details of the
conductivity. In large random networks the relation (\ref{definition})
suffices to define $\Sigma_{ij}$ without need of a detailed
specification of boundary conditions. However, out {\em exact}\/
result is valid also for small networks of arbitrary topology,
provided that the electric field $\vec{E}$ is generated using the very
specific boundary conditions specified in the Appendix.


\end{multicols}


\begin{thebibliography}{99}
\bibitem{remark}
In many textbooks the shear modulus is defined by $\mu=C_{44}$. This
inaccurate definition is relevant to regular solid for which, under
standard temperature and pressure conditions, $C_{44}\gg P$. We
show in this section that for Gaussian networks $C_{44}=0$, and
therefore they represent the extreme opposite case.
\end{thebibliography}
\end{document}